\documentclass[aps,twocolumn]{revtex4-1}
\usepackage{epsfig}
\usepackage{color}
\usepackage{amsmath}
\usepackage[retainorgcmds]{IEEEtrantools}
\usepackage{ulem}

\begin{document}

\title{Fundamental ingredients for the emergence of 
discontinuous phase  transitions in the  majority vote model}

\author{Jesus M. Encinas, Pedro E. Harunari$^1$, M. M. de Oliveira$^2$ and 
C. E. Fiore$^1$}
\email{fiore@if.usp.br} 

\affiliation{$^1$ Instituto de F\'{\i}sica,
Universidade de S\~{a}o Paulo, \\
Caixa Postal 66318\\
05315-970 S\~{a}o Paulo, S\~{a}o Paulo, Brazil\\
$^2$Departamento de F\'{\i}sica e Matem\'atica,
CAP, Universidade Federal de S\~ao Jo\~ao del Rei,
Ouro Branco-MG, 36420-000 Brazil. }
\date{\today}

\begin{abstract}
Discontinuous transitions have received  considerable interest
 due to the uncovering that many phenomena such as catastrophic changes, 
epidemic outbreaks and synchronization   present a behavior signed by abrupt (macroscopic) changes (instead
of smooth ones) as a tuning parameter is changed. However, in different cases there are still scarce  microscopic models 
reproducing  such above trademarks.
With these ideas in mind, we investigate the fundamental ingredients
underpinning the discontinuous transition in one
of the simplest systems with up-down $Z_2$ symmetry
recently ascertained in [Phys. Rev. E {\bf 95}, 042304 (2017)].
Such system, in the presence of an extra ingredient-the inertia- has its
continuous transition  being switched to a discontinuous one in complex
networks. We  scrutinize the role of three fundamental
ingredients: inertia, system degree, and the lattice topology. Our analysis has been carried out for regular
lattices and random regular networks with different node degrees (interacting neighborhood) through mean-field
treatment and numerical simulations. Our findings reveal that not only the inertia but also the connectivity
constitute essential elements for shifting the phase transition. 
Astoundingly, they also manifest in low-dimensional regular topologies, exposing a scaling behavior entirely
different than those from the complex networks case. Therefore, our findings put on firmer bases the essential
issues for the manifestation of discontinuous transitions in such relevant class of systems with $Z_2$ symmetry.
\end{abstract}
\flushbottom
\maketitle

Spontaneous breaking symmetry manifests in a countless 
sort of systems besides the classical
ferromagnetic-paramagnetic phase transition \cite{marr99,henkel}.
 For example, fishes
moving in ordered schools, as a strategy of protecting
themselves against predators, can suddenly reverse the direction of their motion due to
the emergence of some external factor,
such as water turbulence, or opacity \cite{vicsek}. 
 Also, some species of Asian fireflies 
  start (at night) emitting unsynchronized flashes of light but, 
some time later, the whole swarm is flashing in a coherent way \cite{aceb}.
In social systems as well, order-disorder transitions describe
the spontaneous formation of a common language, culture or 
the emergence of consensus \cite{loreto}. 

Systems with $Z_2$ (``up-down") symmetry
constitute ubiquitous models of spontaneous breaking symmetry, and their phase transitions
and universality classes have 
been an active topic of research during the last 
decades \cite{marr99,odor07,henkel}.
Nonetheless, several transitions between the distinct regimes 
do not follow smooth behaviors
\cite{hans,coinfect1,paula}, but instead, they manifest through 
abrupt  shifts. These {\it discontinuous} (nonequilibrium) transitions
have received much less attention 
than the critical transitions and a complete understanding of their
fundamental aspects is still lacking. In some   system classes,   essential 
mechanisms for their occurrence \cite{fiore14}, competition with  
distinct dynamics \cite{martins,salete1},  phenomenological finite-size theory \cite{fss15}
and others \cite{dic01,weron,scp1,temp} have been pinpointed.

Heuristically, the occurrence of a continuous transition in 
  systems with $Z_2$ symmetry is described (at a mean field level)
by the logistic equation
$\frac{d}{dt}m=am-bm^3$, that exhibit the steady solutions $m=0$
and $m=\pm \sqrt{a/b}$. The first solution is stable for negative values
of the tuning parameter $a$, while the second is stable for positive values of $a$. For the description
of abrupt shifts, on the other hand, one requires the inclusion
of an additional term $+cm^5$, where $c>0$ ensures finite values
of $m$. In such case, the jump of $m$ yields at $a=\frac{b^2}{4c}$, reading 
$\pm \sqrt{b/2c}$. Despite portrayed
  under the simple above logistic equation, there
are scarce (nonequilibrium) 
{\it microscopic} models forecasting discontinuous transitions. 

Recently, Chen et al. \cite{chen2} showed that the usual majority
vote (MV) model, an emblematic example of nonequilibrium system
with $Z_2$ symmetry \cite{mario92,chen1,pereira}, exhibits a discontinuous
 transition in complex networks, provided relevant strengths of inertia (dependence on the local spin)
  is incorporated in the dynamics.  This results in 
a stark contrast with the  original (non-inertial) MV, 
whose phase transition is second-order, irrespective the lattice 
topology and neighborhood.   The importance of such results is highlighted by the fact that behavioral inertia is an essential
characteristic of human being and animal groups. Therefore,
inertia can be a significant ingredient triggering abrupt transitions
that arise in social systems \cite{loreto}.

Although inertia plays a fundamental role 
for changing the nature of the phase transition, their effects allied to
 other components have not been satisfactorily understood yet \cite{harunari}.
 More concretely, 
does the phase transition become 
discontinuous  irrespective of the 
neighborhood or on the contrary, is it required a minimal neighborhood
for (additionally to the inertia) promoting a discontinuous shift?         
Another important question concerns the topology of the network. Is it a fundamental
ingredient? Do complex and low-dimensional regular structures bring us similar
conclusions?

Aimed at addressing questions mentioned above, here we examine
separately, the role of three fundamental ingredients: inertia,  
system degree, and the lattice topology. 
For instance, we consider regular lattice and random regular (RR) networks
for different system degrees through mean-field treatment
and numerical simulations. Our findings point
out that a minimal neighborhood is also an essential  element
for promoting an abrupt  transition. Astonishing, a  discontinuous 
transition is also  observed in low-dimensional regular networks, whose
scaling behavior is entirely different from that
presented in complex networks \cite{fss15}. Therefore, our
upshots  put on firmer bases the  minimum and essential issues 
for the manifestation of ``up-down'' discontinuous transitions.

\section*{Model and results}
In the original MV, with probability $1-f$
each node $i$ tends  to align itself with its local neighborhood majority
 and, with complementary probability $f$, the majority rule is not
followed.  By increasing the misalignment parameter  $f$, 
a continuous order-disorder phase transition takes place, irrespective
the lattice topology \cite{mario92,chen1,pereira}. Chen et al. 
\cite{chen2} included in the original model a
 term proportional to the
local spin $\sigma_i$, with strength $\theta$, 
given by 
\begin{equation}
  w_i(\sigma)=\frac{1}{2}\left\{1-(1-2f)\sigma_{i}
  {\rm sign}\left[(1-\theta)\sum_{j=1}^{k}\sigma_j/k+\theta \sigma_i\right]\right\},
\label{eq2}
\end{equation}
where ${\rm sign}(X)=\pm 1$, according to $X>0$ and $<0$. Note that
one recovers the original rules as $\theta=0$.

\begin{figure}[ht]
\centering
\epsfig{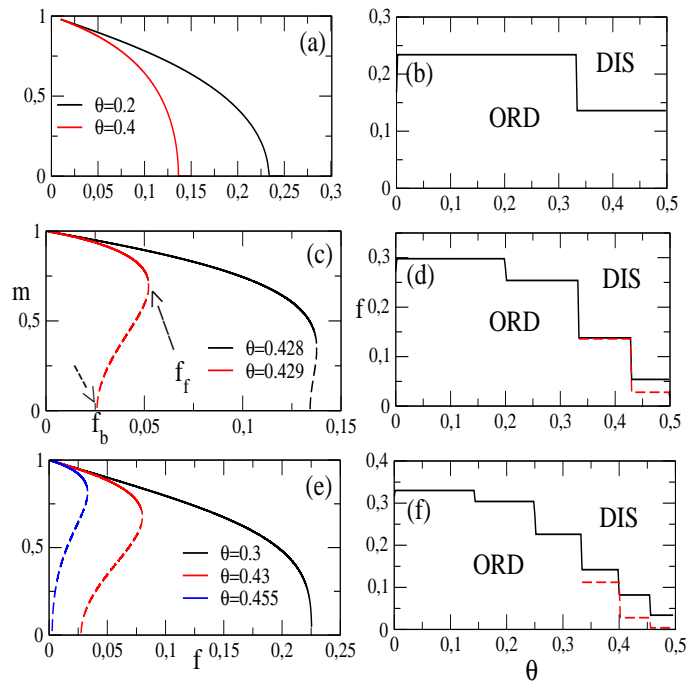}
\caption{From top to bottom,
    mean-field results for regular networks
for $k=4$, $k=8$  and $k=12$, respectively. The left
    panels show the behavior of $m$ versus $f$ for
    distinct $\theta$'s, whereas the right ones show the respective
    phase diagrams.  ORD and DIS correspond to the ordered
    and disordered phases, respectively.
    Location of forward and backward transitions are
    exemplified by arrows in panel $(c)$.}
\label{fig1}
\end{figure}

{\it MFT results}: 
In several cases, a mean field treatment affords a good description of the model properties.
By following the main steps from Refs. \cite{romualdo,chen1,chen2,harunari},
we derive relations for evaluating the order parameter $m$ for
fixed $f, \theta$ and $k$ [see Methods, Eqs. (\ref{eq3})-(\ref{eq8})].
Fig.  \ref{fig1}  shows the main results
for $k=4,8$ and $12$.
Note that MFT predicts a
continuous phase transition for $k=4$  irrespective
the value of $\theta$ [see panels $(a)$ and $(b)$],
in which   $m$ is a decreasing monotonic
function of the misalignment parameter $f$. 
 An opposite scenario is drawn for $k=8$ and $12$, where
phase coexistence stems as  $\theta$ increases [see panels $(c)-(f)$].
They are signed by the presence of a spinodal curve,
emerging at $f_b$ [see e.g panel $(c)$ and $(e)$] 
and meeting the monotonic decreasing branch
at $f_f$. For $k=8$, the coexistence line  arises
only when $\theta >1/3$ and is very tiny ($f_f-f_b$ is about $2.10^{-4}$), 
but they are more pronounced for  $\theta>3/7$. Analogous
phase coexistence hallmarks 
 also appear
for $k=12$ (panel $(e)$) and $k=20$ (Fig. \ref{fig5}  and \cite{chen2}).
 Thus, MFT insights us that
 large $\theta$ and $k$ ($k>6$)  are fundamental ingredients for the appearance
of a discontinuous phase transition. 
A remarkable feature  concerning
the phase diagrams is the existence  of plateaus,  
in which the transition points  present identical
values within a
range of inertia values. As it will be explained
further, that is a consequence of 
 the regular topology. Also, the   number of plateaus 
increase by raising $k$.

\begin{figure}[ht]
\centering
  \epsfig{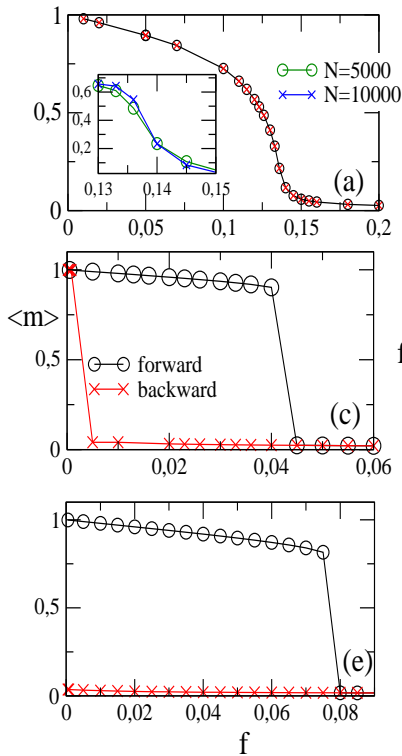}
  \caption{RR Networks: From the top to bottom,
    numerical results for $k=4$, $k=8$  and $k=12$, respectively. The left
    panels exemplify the behavior of $\langle m\rangle$ versus $f$ for
    $\theta=0.33$ ($k=4$) and $0.35$ ($k=8$ and $12$), whereas
    right ones show the phase diagrams. Inset: Reduced
  cumulant $U_4$ vs. $f$ for $\theta=0.2$. Circles (times) correspond
to the increase (decrease) of $f$ starting from an ordered (disordered) phase.}
\label{fig2}
\end{figure}

{\it Numerical results:}
Numerical simulations  furnish more realistic outcomes than the
MFT ones, since the dynamic fluctuations are taken into account.
The actual simulational protocol is described in [Methods].
Starting with the random topology, 
Fig. \ref{fig2} shows the phase diagrams for $k=4,8$, and $k=12$,
respectively. 

First of all, we observe that the positions
of plateaus are identical than those predicted from the MFT. Also, the phase
transition is continuous for $k=4$, irrespective
the inertia value. In all cases (see e.g Fig. \ref{fig2} $(a)$
for $\theta=0.33$), the phase transition
is absent of hysteresis and  
$U_4$ curves cross at $f_c \sim 0.14$ with $U_0=0.23(2)$.
For $\theta>1/3$, no phase transition is displayed and the
system is constrained into the disordered phase. 
Opposite to the low $k$,
discontinuous transitions
are manifested for $k=8$ and $12$ in the regime of pronounced $\theta$. More
specifically, the crossovers take place
at $\theta=1/3$ and $\theta=1/4$ for the former and latter $k$,
respectively.
Notwithstanding, there are some differences between
approaches. As expected, MFT predicts
overestimated transition points than numerical simulations.
Although MFT predicts a continuous phase transition
in the interval  
$\frac{1}{4}<\theta<\frac{1}{3}$ ($k=12$),
numerical simulations suggest that it is actually discontinuous ones.

\begin{figure}[ht]
\centering 
\epsfig{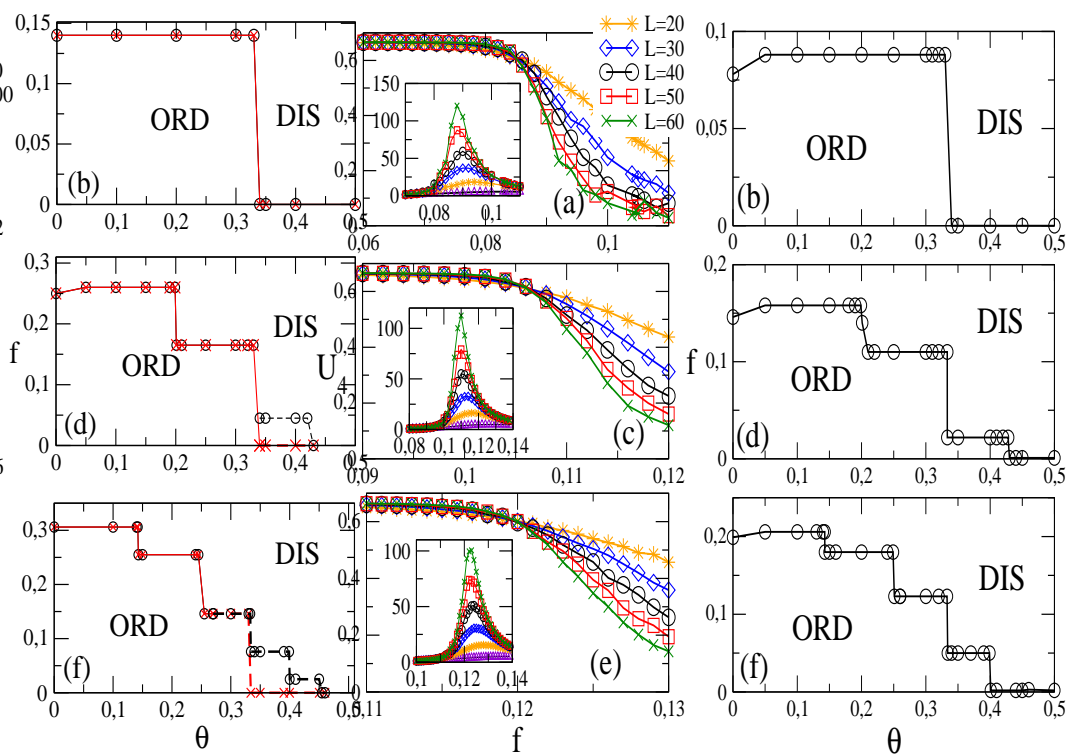}
\caption{Bidimensional regular lattices for distinct
system sizes $N=L\times L$: 
Left panels show the reduced cumulant $U_4$ vs $f$
for the nearest neighbor $(a)$, second-neighbor 
$(c)$ and third-neighbor $(e)$ versions,  respectively.
Inset: the same but for the variance $\chi$. 
Right panels show their correspondent phase diagrams. In all cases,
continuous lines correspond to critical phase transitions.}
\label{fig3}
\end{figure}

\begin{figure}[ht]
\centering
  \epsfig{file=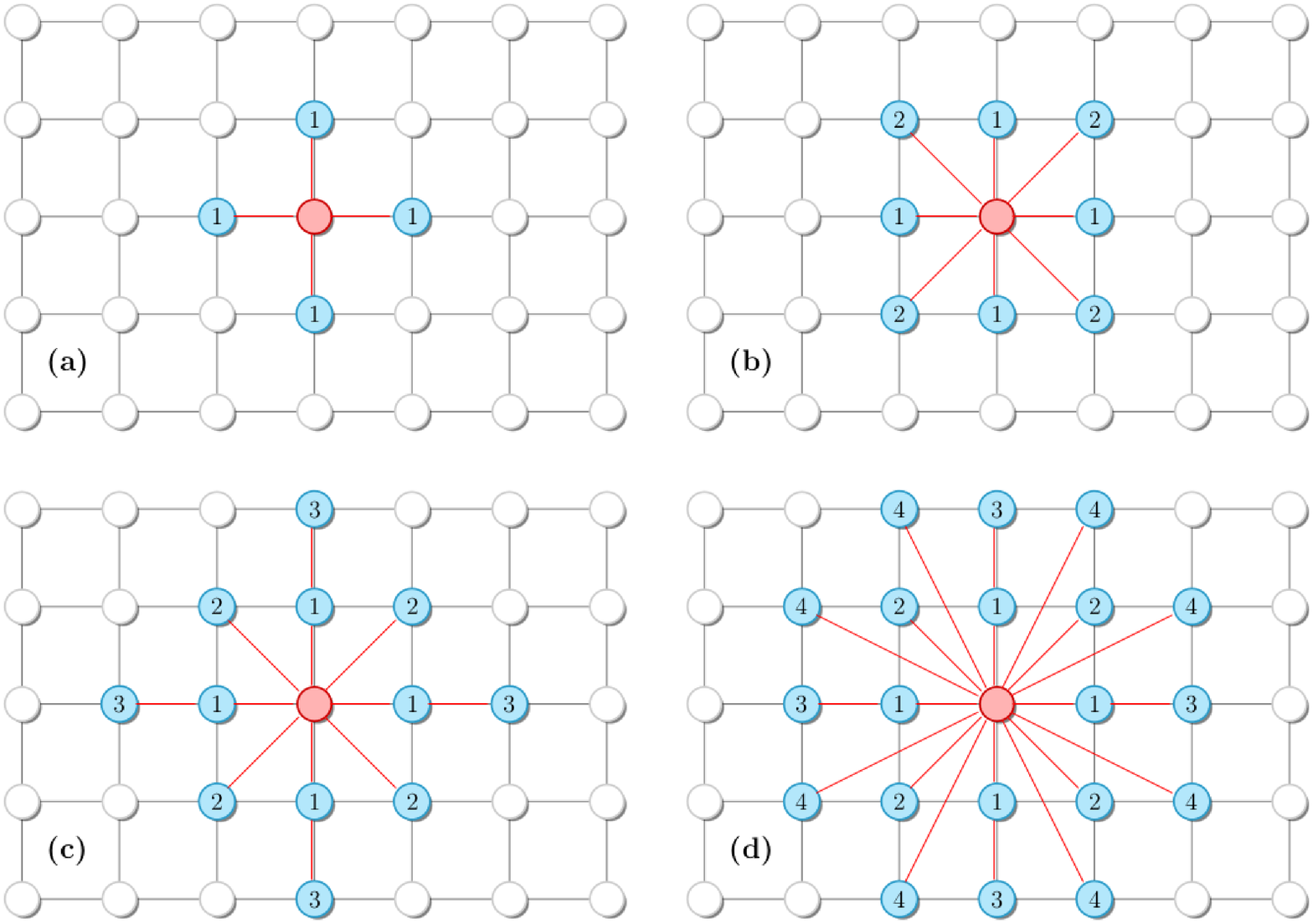,width=8cm,height=6cm}
  \caption{Local configuration for the versions
     with interactions between the first $(a)$,
    first and second $(b)$, first to third $(c)$
  and first to fourth $(d)$ next neighbors.}
\label{fig0-1}
\end{figure}

In Fig. \ref{fig3}, the previous  analysis is extended for
regular (bidimensional) versions. In order
to mimic the increase of connectivity, the cases  $k=4, 8$ and $12$
cases are undertaken by restricting the interaction
between the first, first and second, 
first to third next neighbors, as exemplified
in panels $(a)-(c)$ in Fig. \ref{fig0-1}, respectively.

The position of plateaus are identical than
both previous cases, but with lower $f_c$'s. This is roughly understood
 by recalling that  homogeneous complex networks
exhibit mean-field structure, whose
correspondent transition points are thus
larger than those from regular lattices.
Similarly, all critical points
are obtained from  the crossing among $U_4$ curves,
but the value $U_0^{*}$ 
is different from the RR case, 
following to the Ising universality
class  value $U_0^{*} \sim 0.61$ \cite{mario92,marr99,odor07}.
Thereby, there is an important difference
between random and regular structures: 
The phase transitions are continuous  irrespective
the inertia value for $k$ from $k=4$ to $k=12$. 

\begin{figure}[ht]
\centering
\epsfig{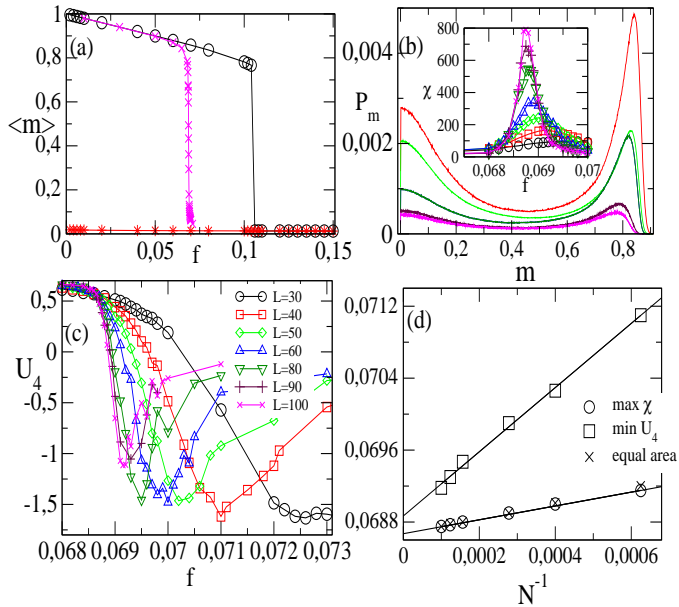}
\caption{Results for $k=20$ and $\theta=0.35$:
  Panel $(a)$ compares the order parameter $\langle m \rangle$ versus
  $f$ for the RR network (circles and stars) and regular lattice
  (symbol $\times$).  
  Regular lattice case: Panels $(b)$ and $(c)$
  show  the equal area probability distribution 
  and  the $U_4 \times f$ for distinct $L$'s ($N=L\times L$), respectively. Inset:
  The variance $\chi$ versus $f$. In $(d)$, the positions of maxima  of $\chi$,
  minima of $U_4$ and equal area versus  $1/N$.}
\label{fig4}
\end{figure}

\begin{figure}[ht]
\centering
\epsfig{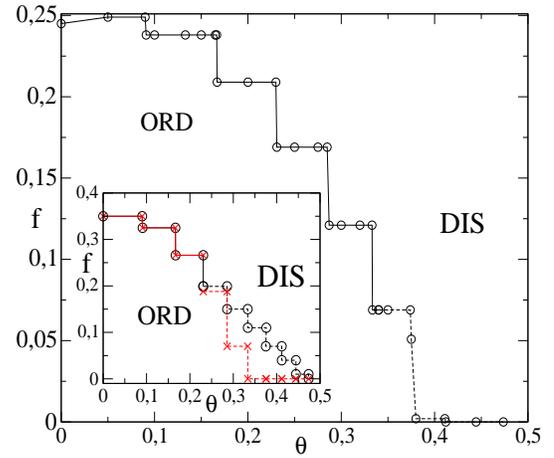}
\caption{The phase diagram $\theta$ versus $f$ for
the MV with $k=20$ in a bidimensional lattice. Continuous
and dashed lines correspond to critical and discontinuous
phase transitions, respectively. Inset: The same, but
for the RR topology. 
Circles (times) correspond
to the increase (decrease) of $f$ starting from an ordered (disordered) phase. }
\label{fig5}
\end{figure}

An entirely different scenario is unveiled
by  extending interactions range
up to the fourth next neighbors spins (mimicking the case $k=20$) and large inertia values
[see e.g  Fig. \ref{fig0-1}$(d)$], in which
 the phase transition
becomes discontinuous 
(see e.g. Fig. \ref{fig4} for $\theta=0.35$).
Contrary to the random complex case, hysteresis
is absent [panel $(a)$] and the order-parameter distribution
exhibits a bimodal shape [panel $(b)$].   Complementary, $U_4$ 
presents a minimum whose value decreases with $N$  [panel $(c)$] and the maximum
of $\chi$ increases with $N$  (inset). In all cases,
the $f_N$'s (estimated from que equal
area position, maximum of $\chi$ and minimum
of $U_4$) scales  with $N^{-1}$  [panel $(d)$], in consistency with Ref. \cite{fss15},
from which one obtains the estimates $f_0=0.0687(1)$ (equal
area and maximum of $\chi$) and $f_0=0.0689(1)$ (minimum
of $U_4$) [see Methods for obtaining the finite-size scaling relation].

 In Fig. \ref{fig5}, the phase diagram is presented.
As in previous cases, the positions of the plateaus are identical to
the $RR$ for $k=20$ (see inset and Ref. \cite{chen2}). The phase
coexistence occurs for  $\theta>1/3$,  larger than 
$\theta>3/13$ (RR structure). For $\theta<1/3$, the phase transition is continuous, 
although $U_4$ presents a value
different from $U_0^{*} \sim 0.61$ in the interval $2/7<\theta<1/3$.

\subsection*{Origin of plateaus}

Since the transition rate depends only on the signal of resulting argument
in Eq. (\ref{eq2}), the phase diagrams will present plateaus provided
the number of neighbors is held fixed.
Generically, let us take a lattice of degree
  $k$ with the central
site $\sigma_0$ with $n_{k}^{+}$ and $n_{k}^{-}$ nearest neighbors
with spins $+1$ and $-1$, respectively (obviously
$n_{k}^{+}+n_{k}^{-}=k$). Taking for instance  $\sigma_0=-1$ 
(similar conclusions are earned for $\sigma_0=1$). In such case, the 
 argument of ${\rm sign(X)}$ reads 
$1-\frac{2n_{k}^{-}}{k}-2\theta(1-\frac{n_{k}^{-}}{k})$, implying
that for all  $\theta<\theta_{p}=\frac{k-2n_{k}^{-}}{2(k-n_{k}^{-})}$ the 
transition 
rate $-1\rightarrow +1$ will be performed with the same rate $1-f$
and thus the  transition points are equal. 
Only for $\theta > \theta_p$
the transition $-1 \rightarrow +1$ is performed with  probability $f$. 
Table I lists the plateau points
$\theta_p$ for $k=8$ and distinct $n_{k}^{-}$'s.  
For example, for $n_k^{-}=3$ and $0< \theta<\theta_p=\frac{1}{5}$, 
all transition rates are equal, implying  the same $f_c$
for such above set of inertia.
For $\theta=\theta_p=\frac{1}{5}$ the second local configuration
becomes different and thereby $f_c$ is different
from the value for $\theta<\theta_p$.  Keeping so on with other values
of $n_{k}^{-}$,  the next plateau positions are located. It is worth mentioning
that  $n_{k}^{-}>n_{k}^{+}$ leads to negative $\theta_p$'s, 
that not have been examined here.
\begin{table}
\centering
  \begin{tabular}{|c|c|c|c|c|c|} \hline
    \text{$\sigma_0$} & \text{$n_{k}^{+}$} & \text{$n_{k}^{-}$}  & $\text{X}>0$ &
    \text{$\theta_p$} \\ \hline \hline
-1 & \text{4+} & \text{4-} & \text{-$\theta>0$} & \text{0} \\ \hline
-1 & \text{5+} & \text{3-}  & $\frac{1-5\theta}{4} >0$ & $\frac{1}{5}$ \\ \hline
-1 & \text{6+} & \text{2-}  & $\frac{1-3\theta }{2} >0$ & $\frac{1}{3}$ \\ \hline
 -1 & \text{7+} & \text{1-}  & $\frac{3-7\theta }{4} >0$ & $\frac{3}{7}$ \\ \hline
 -1 & \text{8+} & \text{0-} & $1-2 \theta >0$ & $\frac{1}{2}$ \\ \hline
\end{tabular}
  \caption{For the central site $\sigma_0=-1$
    and connectivity $k=8$, the signal function
    for distinct local configurations. $X$ is the value of 
    resulting expression
    $1-\frac{2n_{k}^{-}}{k}-2\theta(1-\frac{n_{k}^{-}}{k})$ (see main Text)
  and $\theta_p$ denotes the plateaus positions.}
\end{table}

\section*{Methods}

We  consider a class of systems in which
each site $i$  can take only two values 
$\pm 1$, according to its ``local spin'' (opinion)  $\sigma_i$, is
``up'' or ``down'', respectively. The time evolution of the probability
$P(\sigma)$
associated to a local configuration $\sigma \equiv (\sigma_1,..,\sigma_i,\sigma_N)$
is ruled by the master equation
\begin{equation}
\frac{d}{dt}P(\sigma,t)=\sum_{i=1}^N\{w_{i}(\sigma^i)P(\sigma^i,t)-
w_{i}(\sigma)P(\sigma,t)\},
\end{equation}
where the sum runs over the $N$ sites of the system and $\sigma^i \equiv (\sigma_1,..,-\sigma_i,\sigma_N)$
differs from $\sigma$ by the local spin of the $i-$th site.
From the above, the time evolution of  the  magnetization  of
a  local site, 
defined by $m=\langle \sigma_i \rangle$, is given by 
\begin{equation}
\frac{d }{dt}m=\left(1-m\right)w_{-1 \rightarrow 1}-
\left(1+m\right) w_{1 \rightarrow -1},
\label{eq3}
\end{equation}
where $w_{-1 \rightarrow 1}$ and $w_{1 \rightarrow -1}$ denote the transition
rates to  states with opposite spin.  In the steady state, one has that
\begin{equation}
m=\frac{w_{-1 \rightarrow 1}-w_{1 \rightarrow -1}}{w_{-1 \rightarrow 1}+
w_{1 \rightarrow -1}}.
\label{eq4}
\end{equation}

By following the formalism  from Refs. \cite{chen2,romualdo,harunari},  
the transition rates $w_{-1 \rightarrow 1}$ and $w_{1 \rightarrow -1}$ 
in Eq. (\ref{eq4}) are  decomposed as
\begin{equation}
w_{-1 \rightarrow 1}=(1-2f){\bar P_{-}}+f,
\end{equation}
and
\begin{equation}
w_{1 \rightarrow -1}=(1-f)-(1-2f){\bar P_{+}},
\end{equation}
where  ${\bar P_{-}}$(${\bar P_{+}}$) denote
 the probabilities that the node $i$ of 
degree $k$, with spin
$\sigma_i=-1$ ($\sigma_i=1$) changes its state according to the 
majority (minority) rules, respectively. Such probabilities  can be written
according to
\begin{equation}
{\bar P_{\pm}}=\sum_{n=\lceil n_k^{\pm} \rceil}^{k}(1-\frac{1}{2}\delta_{n,n_{k}^{\pm}})C_{n}^{k}
p_{+1}^{n}p_{-1}^{k-n},
\label{eq5}
\end{equation}  
with $p_{\pm 1}$ being the probability that a nearest neighbor is
$\pm 1$ and $n_{k}^{-}$ and $n_{k}^{+}$  corresponding to the lower limit 
of the ceiling function, reading $n_{k}^{-}=\frac{k}{2(1-\theta)}$
and $n_{k}^{+}=\frac{k(1-2\theta)}{2(1-\theta)}$.

 Since we are dealing with uncorrelated structures with 
the same degree $k$, $p_{\pm}$ is simply 
$(1\pm m)/2$, 
from which Eq. (\ref{eq4}) reads
\begin{equation}
\frac{1+m}{2}=
\frac{(1-2f) {\bar P_-}+f}{1+(1-2f)({\bar P_-}-{\bar P_+})},
\label{eq8}
\end{equation}
with ${\bar P_\pm}$ being evaluated from Eq. (\ref{eq5}). 
Thus,  the solution(s) of Eq. (\ref{eq8}) grant
 the steady values of $m$. 

 An alternative way of deriving the MFT expressions consists in
 writing down the 
transition rates  as the sum of products of the local spins 
$w_i(\sigma)=\frac{1}{2}(1-\sigma_i\sum_{A}c_A\sigma_A)$,where
$\sigma_A$ is the product of  spins belonging 
to the cluster of $k$ sites, and $c_A$ is a real coefficient. 
For example, for $k=8$ and $\theta=0$, we have 
that $\frac{d}{dt}m=-m+(1-2f)\{\frac{35}{16}m-\frac{35}{16}m^3+\frac{21}{16}m^5-\frac{5}{16}m^7\}$,
yielding the critical point  $f_c=\frac{19}{70}$, in full equivalency with $f_c$ 
obtained from Eq. (\ref{eq8}).

\begin{figure}[ht]
\centering
\epsfig{file=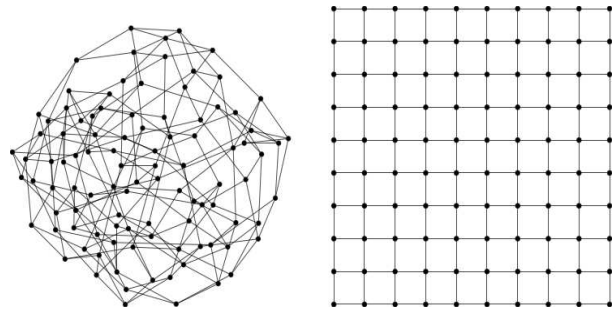,width=8cm,height=4cm}
\caption{Examples of systems with $N=100$ sites and neighborhood $k=4$:
regular random network (left) and regular lattice (right).}
\label{fig0}
\end{figure}

The numerical simulations will be grouped into two parts: Random regular (RR) network and 
(low dimensional) bidimensional lattices.
In the former structure,
each site $i$ (also referred as node or vertex) 
is linked, at random, to $k$ neighbors. 
In the latter, the neighborhood is also $k$, but they
form a regular arrangement. Note that both structures are quenched, i.e., 
they do not change during the simulation of the model. 
Fig. \ref{fig0}  exemplifies both structures, 
for a system with  $100$ sites and connectivity $k=4$. 
Periodic boundary conditions have been adopted in the bidimensional
case.

For a given network topology, and with $N$, $f$, and $\theta$ held fixed, a site $i$ is randomly chosen,
and its spin value $\sigma_i$ is updated 
($\sigma_i \rightarrow -\sigma_i$) according to Eq. (\ref{eq2}).  With
complementary probability, the local spin remains unchanged.
A Monte Carlo (MC) step corresponds to $N$ updating spin trials. 
After repeating the above dynamics a sufficient
number of MC steps, the system attains a nonequilibrium steady state. Then,
 appropriate quantities, including the mean magnetization 
 $\langle m \rangle=\frac{1}{N}\langle| \sum_{i=1}^{N}\sigma_i|\rangle$, 
its variance 
$\chi=N[\langle m^{2} \rangle-\langle m \rangle^{2}]$
 and the fourth-order reduced cumulant $U_4=1-\frac{\langle m^{4} \rangle}{3\langle m^{2} \rangle^{2}}$ are evaluated, in order to locate the  transition
 point and to classify the phase transition.
 
  A continuous phase transition is trademarked by the algebraic
 behaviors of  $\langle m \rangle \sim N^{-\beta/\nu}$
 and $\chi \sim N^{\gamma/\nu}$, where $\beta/\nu$ and
 $\gamma/\nu$  are their associated critical exponents. Another principal
 feature of continuous transitions is that
 $U_4$, evaluated for distinct $N$'s, intersect 
 at  $(f,U)=(f_c,U_0^{*})$. Although  $U_0^{*}$  and the critical
 exponents depend on the lattice topology \cite{mario92,pereira},
they behave similarly for
 random and regular topologies.
 Off the critical point, $U_4$ reads $U_4 \rightarrow 2/3$ and $0$
for the ordered and disordered phases, respectively, when $N \rightarrow\infty$.

In similarity with the MFT, numerical
analysis of discontinuous  transitions in complex
networks is commonly identified through the presence of
order parameter hysteresis. 
 Starting from a full ordered phase ($|m|= 1$) 
the system will jump to disordered phase ($|m|=0$) at a 
threshold value $f_f$ when $f$ increases. 
Conversely, if the system evolution starts in the
full disordered phase, decreasing $f$, then the ordered phase 
($|m|\neq 0$) will be reached
at $f_b$. Both ``forward'' and ``backward'' curves are not 
expected to coincide themselves at the phase coexistence. 
 
In contrast to complex structures, the behavior
 of discontinuous transitions is less understood in regular lattices.
Recently, a phenomenological finite-size theory
for discontinuous absorbing phase transitions was proposed \cite{fss15}, in which 
no hysteretic nature is conferred, but instead one observes a 
scaling with the inverse of the  system size $N^{-1}$. Here, we extend it for
$Z_2$ up-down phase transitions.
Such relation can be understood by assuming that close to the coexistence point, 
the order-parameter distribution is (nearly)
composed of  a sum of two independent Gaussians, with each
phase $\sigma$  [$\sigma=o$ (ordered) and $d$ (disordered)] 
described by its order parameter value $m_\sigma$ in such a way that
\begin{equation}
P_{N}(m) = P_{N}^{(o)}(m) + P_{N}^{(d)}(m),
\end{equation}
where each term $P_{N}^{(\sigma)}(m)$ reads
\begin{equation}
P_{N}^{(\sigma)}(m) = \frac{\sqrt{N}}{\sqrt{2 \pi}} \,
\frac{\exp[N \{(\Delta f) m -  (m-m_\sigma)^2/(2 \chi_\sigma)\}]}
     {[F_o(\Delta f;N) + F_d(\Delta f;N)]},
 \label{eq9}
\end{equation}
where  $\chi_\sigma$ is the variance of
the  $\sigma-$gaussian distribution,  $\Delta f=f_N-f_0$ denotes
the ``distance'' to the coexistence point $f_0$ and 
each normalization factor  $F_{o(d)}$  reads 
\begin{equation}
F_{o(d)}(\Delta f;N) = \sqrt{\chi_{o(d)}} \,
\exp\left\{N \Delta f 
\left[m_{o(d)} +  \frac{\chi_{o(d)}}{2} \Delta f
\right]\right\}.
\end{equation}
Note that (\ref{eq9}) leads to the probability distribution
being a sum of two Dirac delta functions centered at $m=m_{o}$
and $m=m_{d}$ at $f=f_0$ for $N \rightarrow \infty$. For
$f-f_0 \rightarrow 0_{+(-)}$, one has a single Dirac delta peak
at $m=m_{d}(m_{o} \neq 0)$.
The pseudo-transition points  can be estimated under
different ways, such as the value of $f_N$ in which
both phases present equal weight (areas).
In such case, from Eq. (\ref{eq9}) it follows that $P_{N}^{(o)}(m)=P_{N}^{(d)}(m)$
for
\begin{equation}
 (f_N - f_0)\, [(m_o - m_d) + \frac{(\chi_o - \chi_d)}{2} 
(f_N- f_0)] =  
\frac{\ln[\chi_d/\chi_o]}{2} \frac{1}{N}.
\label{eq:basic}
\end{equation}
Since $N$ is supposed to be large, the right side of Eq. (\ref{eq:basic}) 
becomes small and thus  $(f_N - f_0)$ is also small. By
neglecting terms of superior order  
$(f_N - f_0)^2$, we have that 
\begin{equation}
f_N  \approx  f_0 +
\frac{\ln[\chi_d/\chi_o]}{2 (m_o - m_d)} \frac{1}{N},
\label{eq:basicsol}
\end{equation}
implying that the difference $f_N-f_0$ scales with the inverse
of the system size $N$. Evaluation of 
the position of peak of variance $\chi$ provides the same dependence
on $N^{-1}$, whose slope is the same that Eq. (\ref{eq:basicsol}) 
[see e.g panel $(d)$ in Fig. \ref{fig4}].
\section*{Conclusions}
A discontinuous phase transition in the standard
majority vote model has been recently discovered
in the presence of an extra ingredient: the inertia.
Results for distinct network topologies revealed the robustness of
such phase coexistence trademarked by hysteresis, bimodal probability
distribution and others features \cite{chen2}.
Here, we advanced by tackling the essential ingredients
for its occurrence. A fundamental conclusion has been ascertained: 
discontinuous transitions  in the MV also manifest in low dimensional regular topologies.
Also, its finite size behavior (entirely different from the network cases),  
is identical to that exhibited by discontinuous phase transitions into absorbing states
 \cite{fss15}. This suggests the existence
of a common and general behavior for first-order transitions in regular
structures. In addition, low connectivity leads to the
suppression of the phase coexistence, insighting us that not only
the inertia is a fundamental ingredient, but also the 
connectivity. For random regular networks, we found that a minimum 
neighborhood is $k=7$, whereas about $k=20$ are required
for changing the order of transition in bidimensional lattices.
Summing up, the present contribution aimed not only stemming the key ingredients for the
emergence of discontinuous transitions in an arbitrary structure, but
also put on firmer basis their scaling behavior in regular topologies.

\section*{Acknowledgements}
We acknowledge the brazilian agencies CNPq, CAPES and  FAPESP for the financial support.

\end{document}